\documentclass[aps,amsfonts,preprint,nofootinbib,superscriptaddress]{revtex4}

\def\S2{\bar{S}}
\def\a{\alpha}
\def\b{\beta}

\def\m{\mu}

\def\t{\theta}

\def\and{a_{n}^\dagger}

\def\sn2d{\Sn2^\dagger}

\def\({\left(}
\def\){\right)}
\def\<{\left\langle}
\def\>{\right\rangle}

\newcommand\ee{\end{eqnarray}}      
\newcommand\be{\begin{eqnarray}}
\newcommand\ba{\begin{array}}           
\newcommand\ea{\end{array}}
\newcommand\eeq{\end{equation}}     
\newcommand\beq{\begin{equation}}

\usepackage{caption}
\usepackage{subcaption}
\usepackage{graphicx}

\begin{document}

\title{Time-dependent Entanglement Entropy in dissipative conformal theories: TFD approach}

\author{M. Dias}
\email{mfedias@gmail.com}
\affiliation{Universidade Federal de S\~ao Paulo, Departamento de F\'isica, Rua S\~ao Nicolau 210, CEP: 09913-030, Diadema, SP, Brasil}

\author{Daniel L. Nedel}
\email{daniel.nedel@unila.edu.br}
\affiliation{Universidade Federal da Integra\c{c}\~{a}o Latino-Americana, Instituto Latino-Americano de Ci\^{e}ncias da Vida e da Natureza, Av. Tancredo Neves 6731 bloco 06, CEP: 85867-970, Foz do Igua\c{c}u, PR, Brasil}

\author{C. R. Senise Jr.}
\email{carlos.senise@unifesp.br}
\affiliation{Universidade Federal de S\~ao Paulo, Departamento de F\'isica, Rua S\~ao Nicolau 210, CEP: 09913-030, Diadema, SP, Brasil}

\begin{abstract}
In this work the TFD formalism is explored in order to study a dissipative time-dependent thermal vacuum. This state is a consequence of a particular interaction between two theories, which can be interpreted as two conformal theories defined at the two asymptotic boundaries of an AdS black hole. The initial state is prepared to be the equilibrium TFD thermal vacuum. The interaction causes dissipation from the point of view of observers who measure observables in one of the boundaries. We show that the vacuum evolves as an entangled state at finite temperature and the dissipative dynamics is controlled by the time-dependent entropy operator, defined in the non-equilibrium TFD framework. We use lattice field theory techniques to calculate the non-equilibrium thermodynamic entropy and the finite temperature entanglement entropy. We show that both grow linearly with time. 
\end{abstract}
\maketitle


\section{Introduction}

Entanglement entropy has been playing an important role in several areas of theoretical physics. From condensed matter where, for some spin  and topological systems, entanglement provides a useful order parameter \cite{Kitaev, Levin}, up to high energy physics, where entanglement plays an important role in the holographic principle, helping to understand how to storage quantum data into classical spacetime \cite{RT}.

An important task in many-body physics is to distinguish intrinsic quantum entanglement from the thermal fluctuations included in the definition of the ordinary quantum entanglement at finite temperatures. A method based on Thermo Field Dynamics (TFD) was developed in \cite{masuo}, which allows to perform such endeavour. In the present work, we will explore a similar method for studying entanglement entropy in dissipative systems. The core of the TFD formalism is the doubling of the degrees of freedom, defining an auxiliary Hilbert space, and a Bogoliubov transformation to entangle such duplicated degrees. The Bogoliubov transformation defines a thermal vacuum such that the statistical average of an operator can be written as the expectation value in this state.
   
An important ingredient of the TFD formalism is the construction of an operator whose expected value in the thermal vacuum provides the thermodynamic entropy of the system: this is called the entropy operator. Besides, it was shown that this operator can lead a system from zero to a finite temperature. Furthermore,  the entropy operator plays an important role in quantum dissipative theories: it is identified with the time evolution controller, driving it through unitary inequivalent states \cite{garvit,ceravi}. By construction, the entropy operator measures the entanglement between the original system and its copy; actually, the thermal vacuum is the state of maximum entanglement.  Whilst in the TFD formalism  this entanglement is just a technical device to deal with systems at finite temperature, in black hole applications it has important physical consequences: in this case, the auxiliary degrees of freedom have physical meaning. In the AdS/CFT context, it helps to interpret the thermodynamic entropy of the black hole in terms of quantum entanglement. 

Recently, an interesting relationship between the TFD entropy operator, dissipation and entanglement has been explored in the holographic context in \cite{nos}, where a conformal dissipative field theory in which the dissipation process is manifestly related to dynamical entanglement was studied. It is also important to note that a close relation between dissipation and entanglement was experimentally studied  in \cite{EntanglementDissipation}, where it was reported an experiment in which dissipation generates a continuous entanglement between two macroscopic objects.
 
The starting point of Ref.~\cite{nos} is to consider a particular coupling between two systems, defined in the two AdS boundaries, that will drive dissipation in one of the boundaries. The interaction is turned on at $t=0$; before this, the vacuum is dual to two disconnected copies of AdS. It was argued that when the interaction is turned on there is formation of a throat that entangles the vacuum dynamically. Although it was shown that in the large $t$ limit it is possible to make an approximation compatible with the large $N$ limit, and the holographic picture can be understood in terms of Vaidya black holes, the scenario is tricky close to $t=0$. This is because a topology change is necessary in order to connect the two separate spaces, and this mechanism is hard to understand using a classical General Relativity solution. In the present work we are going to study a more natural and involved initial condition: we will use the same dissipative Hamiltonian to evolve the system, but the initial state will be prepared to be a thermal state. The vacuum evolution produces a gaussian time-dependent thermal vacuum different from the one commonly used in the literature (mainly in the ADS/CFT applications), for example in \cite{myers}.

Many works involving applications of entanglement entropy in many-body physics focus on entanglement between degrees of freedom associated with spatial regions. That is, the Hilbert space is geometrically partitioned. However, the Hilbert space can be partitioned into several forms. An important issue about momentum entanglement was developed in \cite{Balasu,decoqft,inflation} and possible holographic interpretations of more general entropies are discussed in \cite{Marika}. The entanglement entropy calculated here is also related to a non-spatial partitioned Hilbert space.

In order to regularize the ultraviolet divergence of the entropy and to make clear the different traces that we are going to take, it will  be made use of the lattice field theory technology, and we are going to make a numerical analysis of the time dependent entropy. In the TFD lattice field theory that we are going to work, it is possible to define an extended density matrix, which makes it easy to define the degrees of freedom that we are going to trace over. We will see that the vacuum state in a finite time will be a time-dependent entanglement state at finite temperature. This kind of time-dependent TFD state can be understood as a perturbation of the equilibrium state at $t=0$. In fact, we will show that the entanglement entropy $S_A(t)$, calculated by the expected value of the entropy operator in the vacuum evolved by the dissipative Hamiltonian, is the thermodynamic entropy at $t=0$. The numerical analysis shows that $S_A(t)$ grows linearly with time. In the density matrix language, the expected value of the entropy operator takes into account the trace over the entire auxiliary degrees of freedom. When we take the trace over a part of both the auxiliary and original systems, the $t=0$ entropy becomes the finite temperature entanglement between a subset of the  degrees of freedom and its complement. This entropy also grows linearly with time, but oscillate for large time periods, which indicates saturation.  This behavior resembles the typical conformal field entanglement behavior, when the Hilbert space is geometrically partitioned \cite{calabrese}.

This wok is organized as follows: 
\begin{itemize}
\item Sec.~\ref{TFD}: here we briefly review the TFD formalism and its time-dependent formulation using the so-called Liouville-von Neumann (LvN) approach \cite{Kim,KMMS,Lewis}. This approach is a canonical method that unifies the functional Schr\"{o}dinger equation for the quantum evolution of pure states and the LvN equation for the quantum description of either equilibrium or non-equilibrium mixed states. The different types of entropies calculated throughout the work are presented here.
\item Sec.~\ref{DissCT}: we present the dissipative conformal theory. We define the dissipative time-dependent thermal vacuum state and compare it with the time-dependent thermo vacuum studied, for example, in \cite{myers}. The entropy obtained by tracing over the entire auxiliary system is calculated here.
\item Sec.~\ref{LatDissCT}: we define a lattice dissipative conformal theory.
\item Sec.~\ref{exenen}: we use the lattice technology to calculate the time-dependent finite temperature entanglement entropy.
\item Sec.~\ref{NumRes}: we collect the numerical results we work out. 
\item Sec.~\ref{Conc}: we present our concluding remarks. 
\end{itemize}
 

\section{Brief review of the TFD formalism} \label{TFD}

In this section, we present a revision of the TFD formalism. This short  revision will be important to make clear the differences between the time-dependent thermal state studied in this work and the state defined in \cite{malda-hartman,myers}.

In general, the statistical average $\left\langle O\right\rangle$ of an operator $O$ is defined by the functional $\omega\(O\)=Tr\(\rho O\)$:
\begin{equation}
\left\langle O\right\rangle=\frac{{\mbox Tr}[O e^{-\b H}]}{\omega\(1\)} \ ,
\end{equation}
where $\rho=\frac{e^{-\b H}}{\omega\(1\)}$ and $\omega\(1\)$ is the partition function. 
 
The functional $\omega\(O\)$ is called a state in algebraic statistical quantum field theory and the operators form a $C^{\star}$-algebra \cite{emch,haag}. The functional $\omega$ resembles a vector space, so that the algebra equipped with a particular functional admits a reducible representation on a Hilbert space such as a Fock space. This is the formal basis of the TFD formalism, developed by Takahashi and Umezawa \cite{ume2,ume4,rev2,ume1,kha2,kha3}. The main idea is to describe the thermal state as a pure, rather than a mixed state, and to interpret the statistical average as the expectation value of $O$ in a thermal vacuum:
\begin{equation}
\frac{\omega\(O\)}{\omega\(1\)}=\left\langle 0(\theta )\left|O\right|0(\theta )\right\rangle \ . \label{tr1}
\end{equation}

At the equilibrium, the $\theta$ parameter is a function of the temperature. More generally, the core of TFD is to construct a quantum field theory whose vacuum contains the information about the environment under which the system is subjected. In order to achieve such an endeavor, it is necessary to duplicate the degrees of freedom  by constructing an auxiliary Hilbert space, which is a copy of the original system. Once observed that temperature is introduced as an external parameter, the thermal vacuum appears as a condensate state in the duplicated Fock space composed by the physical space and the auxiliary Hilbert space. In the free limit the thermal vacuum is defined by a Bogoliubov transformation, which actually entangles the system and its copy. The auxiliary Hilbert space is denoted by $\widetilde{{\cal H}}$ and the total Hilbert space is the tensor product of the two spaces, ${\cal H} _{T}={\cal H}\otimes\widetilde{{\cal H}}$, with elements $|\Phi\rangle=|\phi,\widetilde{\phi}\rangle$. From the usual annihilation operators, the vacuum is defined by
\begin{equation}
{A}_{k}\left.\left|0\right\rangle\!\right\rangle={\widetilde{A}_{k}}\left.\left|0\right\rangle \!\right\rangle=0 \ ,
\end{equation}
where $\left.\left|0\right\rangle\!\right\rangle=|0\rangle\otimes|\widetilde{0}\rangle$. 

The original and the tilde systems are related by a mapping called tilde conjugation rules, associated with the application of the Tomita-Takesaki modular operator of the algebraic statistical mechanics \cite{oji,lands}:
\begin{eqnarray}
(A_{i}A_{j})\widetilde{}&=&\widetilde{A}_{i}\widetilde{A}_{j} \ , \nonumber \\
(cA_{i}+A_{j})\widetilde{}&=&c^{\ast}\widetilde{A}_{i}+\widetilde{A}_{j} \ , \nonumber \\
(A_{i}^{\dagger })\widetilde{}&=&(\widetilde{A}_{i})^{\dagger } \ , \nonumber \\
(\widetilde{A}_{i})\widetilde{} &=&A_{i} \ ,  \nonumber \\
\left[\widetilde{A}_{i},A_{j}\right] &=&0 \ .  \label{til}
\end{eqnarray}

The Hamiltonian of the extended system is denoted by $\hat{H}$; it is constructed by demanding that the thermal vacua be the eigenvectors of $\hat{H}$ which are invariant under the tilde conjugation rules, and 
\begin{equation}
\hat{H}\left|0(\theta)\right\rangle=0 \ . \label{hat}
\end{equation}
This can be obtained if we define the following Hamiltonian operator:
\begin{equation}
\hat{H}=H-\tilde{H} \ . \label {Hhat} 
\end{equation}
Note that since the thermal vacuum is annihilated by $\hat{H}$, then in general TFD the tilde system is nonphysical. Indeed, adding the second Fock space corresponds to furnish new thermal degrees of freedom, rather than dynamical degrees of freedom. This is close-banded with the fact that the Hamiltonian $\hat{H}$ is unbounded from below; consequently, the physical observables are defined by operators without tilde.

Due to the sign of $\widetilde{H}$ in the definition of $\hat{H}$, there is an uncountable number of states satisfying Eq. (\ref{hat}), classified by the continuous parameter $\theta$. A choice of  the thermal vacuum is specified by a choice of the thermal Bogoliubov canonical generator $G(\theta)$ such that it commutes with $\hat{H}$. As $G$ changes $\theta$, we have $G\left|0(\theta)\right\rangle\neq0$ and $G$ is a spontaneously broken symmetry. The thermal vacuum is in this manner interpreted as a Goldstone boson.

The first construction of TFD consists in defining a Bogoliubov transformation, using only one generator, that produces an unitary transformation and preserves the tilde conjugation rules. For a set of harmonic oscillators, this generator is given by
\begin{equation}
G(\theta_k)=\sum_k i\theta_k(A_k\widetilde{A}_k-A^{\dagger}\widetilde{A}_k^{\dagger}) \ , \label{bogo}
\end{equation}
and the thermal vacuum is
\begin{eqnarray}
\left|0(\t)\right\rangle&=&\exp(-iG)\left|\theta \right\rangle \nonumber\\
&=& \prod_{k}\frac{1}{\cosh\left(\theta_k\right)}\exp\left[\tanh\left(\theta_k\right)A^{\dagger}_{k}\widetilde{A}^{\dagger}_{k}\right]\left|0 \right\rangle \ . \label{res1a}
\end{eqnarray}
This generator carries out an unitary and canonical transformation such that the creation and annihilation operators transform according to
\begin{eqnarray}
\left(\begin{array}{c}
      A_{k}(\theta) \\
      \tilde{A}^{\dagger}_{k}(\theta)
      \end{array}
\right)&=&e^{-i{\bf G}}\left(\begin{array}{c}
                             A_{k} \\
                             \tilde{A}^{\dagger }_{k}
                             \end{array}
                        \right)e^{i{\bf G}}={\mathbb B}_{k}\left(\begin{array}{c}
                                                                      A_{k} \\
                                                                      \tilde{A}^{\dagger }_{k}
                                                                      \end{array}
                                                                \right) \ ,
\nonumber\\
\left(\begin{array}{cc}
      A^{\dagger}_{k}(\theta) & -\tilde{A}_{k}(\theta)
      \end{array}
\right)&=&\left(\begin{array}{cc}
                A^{\dagger }_{k} & -\tilde{A}_{k}
                \end{array}
          \right){\mathbb B}^{-1}_{k} \ , \label{tbti}
\end{eqnarray}
where the matrix transformation is given by
\begin{eqnarray}
{\mathbb B}_{k}=\left(\begin{array}{cc}
                           \mathfrak{u}_{k} & \mathfrak{v}_{k} \\
                           \mathfrak{v}^{*}_{k} & \mathfrak{u}^{*}_{k}
                           \end{array}
                     \right) \ ,
\qquad|\mathfrak{u}_{k}|^{2}-|\mathfrak{v}_{k}|^{2}=1 \ , \label{tbm1}
\end{eqnarray}
with elements 
\begin{equation}
\mathfrak{u}_{k}=\cosh\left(\theta_k\right) \ ,
\qquad
\mathfrak{v}_{k}=-\sinh\left(\theta_k\right) \ . \label{uvexp}
\end{equation}
 
Since the $A_k(\theta)$ annihilate the thermal vacuum, we have the so called thermal state conditions:
\begin{equation}
(A_k-\tanh(\t_k)\tilde{A}_k)\left|0(\t)\right\rangle=0 \ ,
\end{equation}
and the expectation value of the number operator gives the distribution:
\begin{equation} 
N_k(\t)=\left\langle0(\t)\right|A_k^{\dagger}A_k\left|0(\t)\right\rangle=\sinh^2(\t_k) \ . \label{distri1}
\end{equation}

At the equilibrium, the $\theta$ parameters are fixed as functions of the temperature by minimizing a thermodynamic potential. In order to do this, let us define a very important operator, the entropy operator:
\begin{equation}
K_{A}(\t)=-\sum_{k}\left[A^{\dagger}_{k}A_{k}\ln\sinh^{2}\left(\t_{k}\right)-A_{k}A^{\dagger}_{k}\ln\cosh^{2}\left(\t_{k}\right)\right] \ ,
\end{equation}
such that the thermal vacuum is written as:
\begin{equation}
\left|0(\t)\right\rangle=e^{\frac{K_{A}}{2}}\left|I\right\rangle \ ,
\end{equation}
where 
\beq
\left|I\right\rangle=e^{\displaystyle{\sum _n}A_n^{\dagger}\tilde{A}^{\dagger}_n}\left.\left|0\right\rangle\!\right\rangle \ . \label{estadonn}
\eeq
  
The result would be the same if we had used tilde fields in the definition of the entropy operator. In a thermal equilibrium situation, its expected value provides the thermodynamic entropy. In general, $K_A$ measures the entanglement between the original system and its copy; so, it is a kind of von Neumann entropy operator. To make this statement clear, let us write the vacuum as follows:
\begin{equation}
\left|0(\theta)\right\rangle=\displaystyle{\sum_{n}}\sqrt{{\cal W}_{n}(\theta_n)}\left|n,\tilde{n}\right\rangle \ ,
\end{equation}
where
\begin{equation}
{\cal W}_{n}(\t)=\prod_{k}\left(\frac{\sinh\left(\theta_{k}\right)^{2n_{k}}}{\cosh\left(\theta_{k}\right)^{2n_{k}+2}}\right) \ .
\end{equation}

Now it is possible to make an important connection with the density operator. From this point on we will use a different notation, most commonly used in the AdS/CFT scenario. The two different sets of fields will be denoted by $A$ and $B$ and no longer by $A$ and $\tilde{A}$. Considering the so called extended density matrix
\begin{equation}
\rho=\left|0(\theta)\right\rangle\left\langle0(\theta)\right| \ , \label{extdm}
\end{equation}
if we trace over the tilde fields (now the $B$ fields), the reduced density operator is
\begin{eqnarray}
\rho_{A}&=&Tr_{B}\left[\left|0(\theta)\right\rangle\left\langle0(\theta)\right|\right] \nonumber\\
&=&\sum_{n}{\cal W}_{n}(\theta)\left|n\right\rangle\left\langle n\right| \ ,
\end{eqnarray}
and the entropy is written as
\begin{equation}
{\cal S}_{A}(\theta)=\left\langle 0\left(\theta\right)\left|K _{A}\right|0\left(\theta\right)\right\rangle=\sum_{n}{\cal W}_{n}(\theta)\ln{\cal W}_{n}(\theta) \ . \label{entA-t}
\end{equation}
This makes clear the fact that the entropy operator is the von Neumann entropy relative to the trace over the $B$ degrees of freedom. 

Note that, up to this point, the $\theta$ parameters are quite general and can be time dependent. Now we are going to fix them. In TFD, the expected value of the original Hamiltonian is interpreted as the thermal energy. The following potential is then defined:
\begin{equation}
F=\left\langle 0\left(\theta\right)\left|H\right|0\left(\theta\right)\right\rangle-\frac{1}{\b}\left\langle 0\left(\theta\right)\left|K _{A}\right|0\left(\theta\right)\right\rangle \ .
\end{equation}
Now, for a set of oscillators with frequency $\omega_k$, by minimizing F in relation to $\theta$, we find
\begin{equation}
\sinh^{2}\(\theta_{k}\)=\frac{1}{e^{\beta\omega_k}-1} \ , \label{terdistri}
\end{equation}
which is the usual thermodynamic distribution for a bosonic system in equilibrium at the temperature $1/\beta$. Now the entropy operator provides the thermodynamic entropy and $F$ is the free energy.  We have just shown that in general, in TFD, the same operator that measures the entanglement between two sets of fields provides, for a given choice of the entanglement angle, the thermodynamic entropy. Note that the trace is taken over all the $B$ system. However, the TFD formalism allows one to calculate, in a simple way, a more intricate trace, as shown in \cite{Nakagawa,Dimov,masuo}. Let us explore this fact in the next subsection.


\subsection{The extended entropy} \label{extent}

The Hamiltonian for a system of harmonic oscillators can be written in the number basis as
\begin{equation}\label{HamilMatrixStatic}
\hat H = \sum\limits_{\left\{ {{n_i}} \right\} = 0}^\infty  {\left( {\sum\limits_{i = 1}^N {{\omega_i}\,{n_i} + {\omega_0}} } \right)} \,\left| {{n_1}, \ldots ,{n_N}} \right\rangle \left\langle {{n_1}, \ldots ,{n_N}} \right| \ ,
\end{equation}
where $\{n_i\}=\{n_i\}^N_{i=1}=n_1,\dots,n_N$. In the thermal equilibrium, using (\ref{terdistri}), the extended density matrix (\ref{extdm}) for this system is
\begin{equation}
\rho = \frac{1}{Z}\,\sum\limits_{\{ {n_i}\}  = 0}^\infty  {\sum\limits_{\{ {m_i}\}  = 0}^\infty  {{e^{ - \frac{1 }{2}\, \left( {\sum\limits_{i =1}^N {{\beta\omega_i}\,({n_i} + {m_i}) + 2\,{\beta\omega_0}} } \right)}}} } \,\left| {\{ {n_i}\} } \right\rangle \langle \{ {m_i}\} |\left| {\{ {{\tilde n}_i}\} } \right\rangle \langle \{ {{\tilde m}_i}\} | \ ,
\end{equation}
where $Z$ is the partition function:
\be
Z=\prod\limits_{i=1}^{N}{\frac{{{e^{-\beta \,{\omega_0}}}}}{{1-{e^{-\beta \,{\omega_i}}}}}} \ . 
\ee
If we take the trace of $\rho$ over all the $B$ (tilde) system, we get 
\begin{equation}
 \rho_A=Tr_{B} \ \rho=\sum\limits_{\{ {\ell _i}\}=0}^\infty  {\sum\limits_{\{ {{\tilde \ell }_i}\}  = 0}^\infty  {\langle \{ {\ell _i}\} |\langle \{ {{\tilde \ell }_i}\} |\rho \left| {\left\{ {{\ell _i}} \right\}} \right\rangle |\{ {{\tilde \ell }_i}\} \rangle } } \ .
\end{equation}
In this case we have
\begin{equation}
 \rho _A={\rho }_{eq} = \frac{1}{Z}\,\sum\limits_{\{ {n_i}\}  = 0}^\infty  {{e^{- {\sum\limits_{i = 1}^N {{\beta\omega_i}\,{n_i} - {\beta\omega_0}} }}}} \,\left| {\{ {n_i}\} } \right\rangle \langle \{ {n_i}\} | \ . \label{equilDensityMatrix}
\end{equation}
As expected, we obtained the thermal equilibrium  density matrix. If we calculate the von Neumann entropy associated with this density matrix we obtain the thermodynamic entropy, which is equal to the expected value of the entropy operator. 

Let us now define a different trace. We are going to trace over a subset of oscillators and its tilde correspondent. To do this, let us bipartite the system as follows:
\begin{equation}
\left\{ {{n_i}} \right\}_{i = 1}^N = \left\{ {{n_\mu}} \right\}_{\mu  = 1}^p\bigcup {\left\{ {{n_k}} \right\}_{k = p + 1}^N} \,,\quad p \le N - 1,\quad N \ge 2 \ . \label{parti}
\end{equation}
The extended reduced density matrix $\rho_{AB}$  is obtained as a trace over the parameters of the second system, which now consists of $A$ and $B$ oscillators with index $k$. Note that we are defining two subsets, $A'\subset A$ and $B'\subset B$, which have the same number of degrees of freedom. The idea is to take the trace over the complement of $A'$ and the complement of $B'$ as follows: 
\begin{equation}
 \rho _{AB} = \sum\limits_{\{ {\ell _k}\}  = 0}^\infty  {\sum\limits_{\{ {{\tilde \ell }_k}\}  = 0}^\infty  {\langle \{ {\ell _k}\} |\langle \{ {{\tilde \ell }_k}\} |\hat \rho \left| {\left\{ {{\ell _k}} \right\}} \right\rangle |\{ {{\tilde \ell }_k}\} \rangle } } \ . \label{rhoAB}
\end{equation}
The extended entropy is defined by:
\be
S_{AB}=\rm{Tr}\rho_{AB}\,log\rho_{AB} \ . \label{eentropy}
\ee
In this manner, an easy way to calculate the finite temperature entanglement entropy is achieved.

Establishing the TFD formalism in the thermal equilibrium, it is natural to assume that time-dependent thermal effects can be described by a time-dependent Bogoliubov transformation. In this case, the same form of the  Bogoliubov transformation is used, but the parameters are now time-dependent \cite{umetime}. In general, the Bogoliubov transformed operators are time-dependent and the thermal vacuum is time-independent. We are going to explore the TFD time-dependent formulation using the Liouville-von Neumann (LvN) approach.


\subsection{Time-dependent TFD and the LvN approach} \label{TFDLvN}

In a non-equilibrium situation, a system can interact directly with an environment to make its coupling parameters explicitly time-dependent. In this case, for the harmonic oscillator system, the creation and annihilation operators are now time-dependent and the density matrix $\rho(t)$, built with the time-dependent Hamiltonian  does not satisfy the quantum LvN equation:
\begin{equation}
i\frac{\partial \rho(t)}{\partial t} +[\rho(t),H] =0 \ .
\end{equation}
This makes difficult to relate $1/\beta$ to the equilibrium temperature. One way to solve this problem is given by the LvN approach. The essential idea of the LvN method is that the quantum LvN equation provides all the quantum and statistical information of non-equilibrium systems. The strategy of this approach is to define time-dependent oscillators that satisfy the LvN equation \cite{Kim,KMMS,Lewis}
\begin{equation}
i\frac{\partial a_k}{\partial t} +[a_k,H] =0 \ . \label{LN}
\end{equation} 

The linearity of the LvN equation allows one to use the operators ${a}(t)$ and ${a}^{\dagger}(t)$ to construct operators that also satisfy Eq.~(\ref{LN}), in particular, the number and the density operator. By defining the number operator in the usual way,
\begin{equation}
\hat{N}_k (t) = {a}_k^{\dagger}(t){a}_k(t) \ , \label{ndt}
\end{equation}
one finds the Fock space consisting of the time-dependent number states:
\begin{equation}
\hat{N}_k (t) \vert n_k, t \rangle = n_k \vert n_k, t \rangle \ . \label{numst}
\end{equation}
With these oscillators, a density matrix of the thermal type, which satisfies the LvN equation, can be defined as:
\begin{equation}
\rho_{\rm T}=\frac{1}{Z}\prod_ke^{\beta\omega_0 a_k^{\dagger}(t)a_k(t)} \ , \label{rhoT}
\end{equation}
where $\beta$ and $\omega_0$ are free parameters and $Z$ is the partition function. 

We can use the oscillators ($\ref{LN}$) in the TFD approach and construct the time dependent thermal state:
\be
\left|0(t,\beta)\right\rangle&=&e^{-iG(t)}\left|0,t\right\rangle \ , \nonumber\\
G(t)&=&-i\theta(\beta)\left[a(t)b(t)-a^{\dagger}(t)b^{\dagger}(t)\right] \ .
\ee 
The thermal operators $a(\beta)$ are given by:
\be
a(\beta)=\cosh(\theta)a(t)-\sinh(\theta)b^{\dagger}(t) \ . \label{abt}
\ee
This type of operator appears naturally in the scenario studied here, as will be shown in the following sections.  It is straightforward to define a time-dependent entropy operator in terms of the oscillators that satisfy equation (\ref{LN}). Its expected value becomes
\begin{equation}
S=\displaystyle{\sum_k}\left[(1+ N_{k}(t))\ln(1+N_{k}(t))-N_k(t)\ln N_{k}(t)\right] \ , \label{timeentropy}
\end{equation}
where the time-dependent distribution $N_k(t)$ is the expected value of the number operator defined in (\ref{ndt}). It can be shown that the time-dependent distribution $N_k(t)$ satisfies a Boltzmann equation, as a consequence of a self consistency renormalization procedure \cite{umetime}. This type of time-dependent operator appears naturally in the scenario studied here, as will be shown in the following sections.


\subsection{Holographic interpretation and time-dependent TFD state}

Let us finish this section by presenting the type of contribution that we intend to make with the present work. It is well known that the thermal vacuum of TFD plays an important role in understanding Einstein-Rosen bridges in the ADS/CFT context. When the AdS black hole is written in terms of Kruskal coordinates, the two types of fields appear naturally. Outside the horizon, there are two causally disconnected spacetime geometries (both of which are asymptotically AdS), defining two conformal field theories (CFTs); one of them is interpreted as the degrees of freedom of the auxiliary system. In the Kruskal plane, the solutions of the field equations are uniquely determined by their boundary conditions at the two Minkowski boundaries, in the right ($R$)  and, respectively, left ($L$) quadrants of the Penrose diagram, with reverse clockwise directions, as shown in Fig.~(\ref{fig:fig}) below. Owing to reverse clockwise direction of the $L$ quadrant, while the $A$ operator annihilates outgoing modes at the horizon in the $R$ quadrant, the $\tilde{A}$ operator annihilates ingoing modes in the $L$ quadrant. This is exactly what the tilde conjugation rules make and the reverse clockwise direction is captured in the definition of $\hat{H}$. If the energy levels of the CFT’s are  defined by $E_n$ and the corresponding eigenstates  by $|n\rangle_L, |n\rangle_R$, the entanglement state
\begin{equation}
|\Psi\rangle= \frac{1}{Z^{1/2}}\sum_n e^{\beta E_n/2}|n\rangle_L\times |n\rangle_R \label{dualTFD}
\end{equation}
can be interpreted as the dual description of the eternal ADS black hole at the equilibrium temperature $\beta^{-1}$. Of course, this is also the TFD thermal vacuum for one CFT and it is annihilated by the thermofield hamiltonian $\hat{H}= H_R-H_L$. However, as pointed out in \cite{Maldacena:2013xja,malda-hartman}, the state (\ref{dualTFD}) can also represent two black holes
in disconnected spaces with a common time. The degrees of freedom of the two disconnected spaces do not interact, but the black holes are highly entangled. In this interpretation, the time evolution is upward on both sides and the total Hamiltonian is now $H=H_R+H_L$. The state has a time evolution given by
\begin{figure}
  \centering
    \includegraphics[width=.4\linewidth]{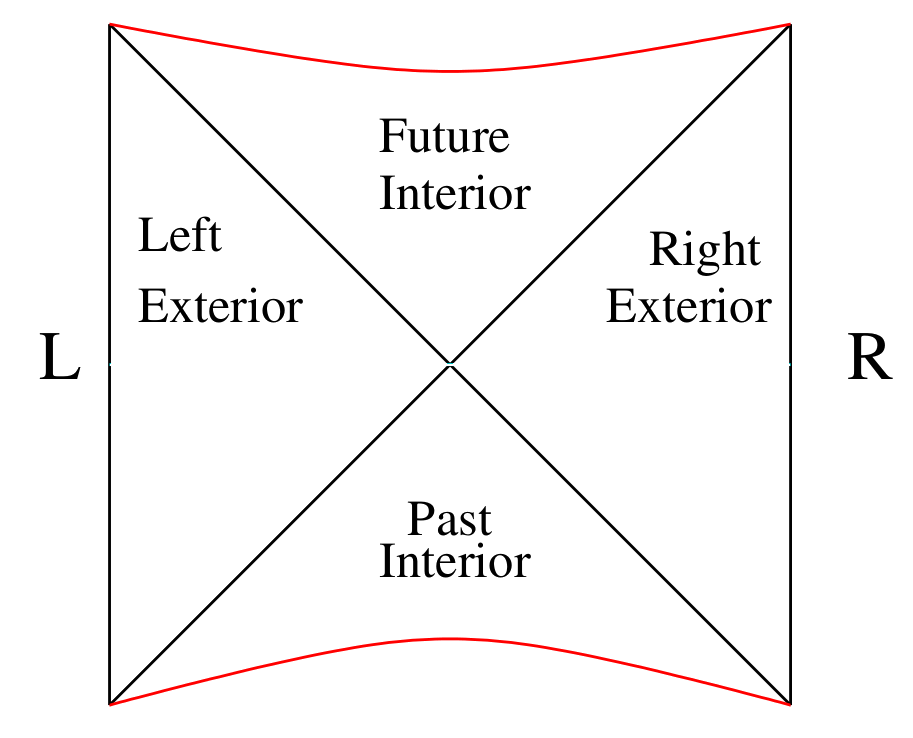}
\caption{Penrose diagram of the eternal ADS black hole. This diagram displays  two exterior regions and
two interior regions.  The right exterior
covers the region that is outside the horizon from the point of view of an observer on
the right boundary. The left exterior is an identical copy and includes a second boundary. The two interior regions contain spacelike singularities.
} \label{fig:fig}
\end{figure}
\begin{equation}
|\Psi(t)\rangle= \frac{1}{Z^{1/2}}\sum_n e^{\beta E_n/2}e^{2iE_n t}|n\rangle_L\times |n\rangle_R \ . \label{dualTFDT}
\end{equation}
Although this state has a clear geometric interpretation, there is no explanation for the Hamiltonian $H$ in the context of the TFD. Actually, in spite of the time evolution of the state (\ref{dualTFD}), the reduced density matrix obtained by tracing out all the $L$ system is exactly equal to (\ref{equilDensityMatrix}). Thus, although this is a time-dependent state, from the point of view of the TFD formalism, this is related to equilibrium thermodynamics.
However, we will show in Sec. V that this is a consequence of taking the trace over all the auxiliary system. We will show that if we take the trace over the complement of the part
of the system and its copy, the resulting density matrix becomes time-dependent.

Here we are going to present an alternative time evolution for the thermal state, more related to the non-equilibrium formulation presented in the previous section. We are going to follow the alternative interpretation of the state ($\ref{dualTFD}$) given in \cite{worm} and \cite{maldanovo}. In this scenario, the state (\ref{dualTFD}) is dual to an  AdS-Schwarzschild wormhole. In this situation, nothing forbids us from coupling the two quantum systems and it was  shown in \cite{worm} that simple couplings between the two sides can render the wormhole traversable. Based on this, we are going to construct a model such that the Hamiltonian is written as $H= H_R-H_L +\Theta(t)H_I$, where $H_I$ contains  interaction terms  between the two sectors and the step function $\Theta(t)$ means that the coupling of the two sides is turned on at $t=0$. As we are going to see in the next section, the interaction between the two sides causes dissipation on one side and the state evolved with the dissipative Hamiltonian is a time-dependent TFD state with different characteristics from the state (\ref{dualTFDT}). In order to explore the Gaussian nature of the state and to perform analytical calculations we will, as in \cite{myers}, analyze the system in the free theory regime, taking into account only the dissipative interaction between the systems. Clearly, in this regime the state is not dual to a black hole, but even so, it has a rich and useful behavior to understand the temporal evolution of entropy.
 

\section{Dissipative conformal theory} \label{DissCT}

A TFD-like entropy operator appears naturally in dissipative theories, as pointed out a long time ago \cite{ceravi}. In this approach, in order to canonically quantize the theory, the degrees of freedom are duplicated and the Hamiltonian is constructed in such a way that the dissipative effects are caused by interactions with the additional (auxiliary) degrees of freedom \cite{ceravi,parentani}.  

Let us start with the following dissipative field equation:
\begin{equation}
(\partial_t^2-\nabla^2)\phi+\gamma\partial_t\phi=0 \ , \label{osci1}
\end{equation}
where $\gamma$ is the damping coefficient. If we introduce a particular interaction of the field $\phi$ with a field $\psi$, this equation can be derived from the following Lagrangian:
\begin{equation}
L=\int_\Omega d^dx\left[\partial_{\mu}\phi\partial^{\mu}\psi+\frac{\gamma}{2}\left(\phi\partial_t\psi-\psi\partial_t\phi\right)\right] \ .
\label{Lagrangeano}
\end{equation}

In \cite{nos}, the AdS/CFT correspondence, applied to AdS black holes, was used to give a holographic interpretation of the dissipative conformal theory defined by Eqs.~(\ref{osci1}) and (\ref{Lagrangeano}). In particular, the ideas developed in \cite{worm} and \cite{maldanovo} were used to connect the two asymptotic boundaries, such that $\gamma$ describes the coupling between the two conformal theories, represented by the fields $\phi$ and $\psi$. In this scenario, the coupling of the two theories defines the dissipative behavior when we study observables defined on one side. The process can be viewed as a teleportation protocol that sets up a quantum coupling of the form $e^{igO_AO_B}$, where $O_A$ and $O_B$ represent the fields living in the two boundaries. The operator $O_A O_B$ can be some linear combination of $a^{\dagger}_{-k},a_k$ and $b^{\dagger}_{k}, b_{-k}$, respectively, where $a$ annihilates $\phi$ modes and $b$ annihilates $\psi$ modes (see \cite{nos} for details). By making the redefinitions
\begin{equation}
\Phi = \frac{\phi+\psi}{\sqrt{2}},\qquad \Psi=\frac{\phi-\psi}{\sqrt{2}} \ ,
\end{equation}
the dissipative Hamiltonian becomes:
\begin{equation}
H = H_{\Phi}-H_{\Psi} \ ,
\end{equation}
where
\begin{eqnarray}
H_{\Phi}&=&\frac{1}{2} \int d^{d}x \left[\left(\Pi_{\Phi}
- \frac{\gamma}{2} \Psi\right)^2 + \frac{1}{2} (\partial_i \Phi)^2\right] \ , \nonumber\\
H_{\Psi}&=&\frac{1}{2} \int d^{d}x \left[\left(\Pi_{\Psi}
+ \frac{\gamma}{2} \Phi\right)^2  + \frac{1}{2} (\partial_i \Psi)^2 \right] \ , \label{hcorrec}
\end{eqnarray}
and $\Pi_{\Phi}$, $\Pi_{\Psi}$ are the canonical momenta conjugated to $\Phi$ and $\Psi$, respectively. In terms of the creation/annihilation operators, the resulting Hamiltonian is 
\begin{equation}
H=H_{0}+H_{int} \ ,
\end{equation}
with
\begin{eqnarray}
H_{0}&=&\sum_{k}\left(\epsilon_k-\frac{\gamma^2}{8\epsilon_k}\right)\left(A^{\dagger}_{k} A_{k}-B^{\dagger}_{k} B_{k}\right) \ , \nonumber\\
H_{int}&=&\sum_{k}\left[-i\frac{\gamma}{2}\left(A_{-k} B_{k}-A^{\dagger}_{-k}B^{\dagger}_{k}\right)+\frac{\gamma^2}{16 \epsilon_k}\left(B_{k}B_{-k}+B^{\dagger}_{k}B^{\dagger}_{-k}-A_{k}A_{-k}-A^{\dagger}_{k}A^{\dagger}_{-k} \right)\right] \ , \label{hamilt} 
\end{eqnarray}
where $A_k$ and $B_k$ correspond to the annihilation operators of the fields $\Phi$ and $\Psi$, respectively, and $\epsilon_k=\sqrt{k^ 2}$. At $t=0$, the vacuum is $\left|0(t=0)\right\rangle =|0\rangle_\Phi\otimes|0\rangle_\Psi$. Since $\left[H_{0},H_{int}\right]=0$, the time evolution of the vacuum state is given by: 
\begin{eqnarray}
\left|0(t)\right\rangle&=&\exp(-iHt)\left|0\right\rangle \nonumber\\
&=&\exp\left[\sum_{k}\; \frac{\gamma t}{2}\left(A_{-k} B_{k}-A^{\dagger}_{-k}B^{\dagger}_k\right)\right]\left|0\right\rangle \nonumber\\
&=&\prod_{k}\frac{\delta_{kk}}{\cosh\left(\frac{\gamma t}{2}\right)}\exp\left[\tanh\left(\frac{\gamma t}{2}\right)A^{\dagger}_{-k}B^{\dagger}_{k}\right]|0\rangle_\Phi\otimes|0\rangle_\psi \ . \label{res1b}
\end{eqnarray}
Note that the dissipative interaction between the two CFTs produces an entangled state. In \cite{nos}, the entanglement entropy was computed and a holographic interpretation in terms of the Vaidya BTZ black hole was given. The entangled state $\left|0(t)\right\rangle$ can be generated by the same entropy operator previously discussed, replacing $\theta$ by $\gamma t$. If the coupling is switched off ($\gamma\to0$), we recover two non-interacting (free) CFTs, whose states describe spacetimes with two locally asymptotic AdS regions \cite{collapse}. In particular, the ground state $|0\rangle_\Phi\otimes|0\rangle_\Psi$ represents two disconnected copies of the exact AdS spacetime \cite{VR}. Here we intend to explore a more involved situation. Note that the limit $\gamma=0$ is the same as the limit $t=0$. Suppose we have a thermal state in $t=0$ and the vacuum is the Kruskal extension of an eternal AdS black hole. The idea is to evolve the thermal vacuum at $t=0$ with the dissipative Hamiltonian (\ref{hamilt}).
 
From now on, we will use the notation $|0\rangle_\Phi\otimes|0\rangle_\Psi=|0\rangle_A\otimes|0\rangle_B$, and write the vacuum at $t=0$ as
\begin {equation}
\left|0(t=0)\right\rangle=\left|0(\beta)\right\rangle=Z^{-1}(\beta)\prod_{k}\exp\left[e^{-E_k\beta/2}A^{\dagger}_{-k}B^{\dagger}_{k}\right]|0\rangle_A\otimes|0\rangle_B \ . \label{TFDstate}
\end{equation}
As discussed in Sec.~\ref{TFD}, this state can be generated by the Bogoliubov transformation defined in (\ref{bogo}). The important point is that the Bogoliubov operator commutes with the Hamiltonian; thus, if we evolve the state $\left|0(t=0)\right\rangle$, we obtain:
\begin{eqnarray}
\left|0(t)\right\rangle&=&\exp(-iHt)\left|0(t=0)\right\rangle \nonumber\\
&=&\exp\left[\sum_{n}\left(\frac{\gamma t}{2}+\theta_n\right)\left(A_{-n}B_{n}-A^{\dagger}_{-n}B^{\dagger}_n\right)\right]|0\rangle_A\otimes|0\rangle_B\nonumber\\
&=&\prod_{n}\frac{\delta_{nn}}{\cosh\left(\frac{\gamma t}{2}+\theta_n\right)}\exp\left[\tanh\left(\frac{\gamma t}{2}+\theta_n\right)A^{\dagger}_{-n}B^{\dagger}_{n}\right]|0\rangle_A\otimes|0\rangle_B \ , \label{res1}
\end{eqnarray}
where
\begin{equation}
\theta_n=\frac{1}{2}\ln\left[\frac{1+e^{-\frac{\beta n}{2}}}{1-e^{-\frac{\beta n}{2}}}\right] \ . \label{thetan}
\end{equation}
The expected value of the number operator is given by
\begin{equation}
N_n(t)=\langle0(t)\left|A_n^{\dagger}A_n\right|0(t)\rangle=\frac{1}{e^{\beta n}-1}\left[\cosh(\gamma t)+e^{\frac{\beta n}{2}}\sinh(\gamma t)\right]^2 \ , \label{timedistri}
\end{equation}
which can be written in the compact form
\begin{equation}
N_n(t)=\langle0(t)\left|A_n^{\dagger}A_n\right|0(t)\rangle=\sinh^2(\theta_n+\gamma t) \ , \label{timedis}
\end{equation}
such that at the $t=0$ or at the $\gamma=0$ limit we have an equilibrium thermal distribution. 
The state $\left|0(t)\right\rangle$ can also be written as 
\begin{equation}
\left|0(t)\right\rangle=e^{\frac{K_A(t)}{2}}\left|I\right\rangle \ ,
\end{equation}
where $\left|I\right\rangle$ is defined in (\ref{estadonn}) and the entropy operator $K_A(t)$ is given by:
\begin{equation}
K_{A}(t)=-\sum_{k}\left[A^{\dagger}_{k}A_{k}\ln\sinh^{2}\left(\t_{k}+\gamma t\right)-A_{k}A^{\dagger}_{k}\ln\cosh^{2}\left(\t_{k}+\gamma t\right)\right] \ .
\end{equation}
Note that the entropy operator carries all the temporal and the thermal dependence of the state. In particular, we can verify that the time evolution of the vacuum state is generated by the time derivative of the entropy operator, or entropy production, namely:
\begin{equation}
\frac{\partial\left|0(t)\right\rangle}{\partial t}=-\frac{1}{2}\frac{\partial K_A}{\partial t}\left|0(t)\right\rangle \ . \label{eq-mov-entropy}
\end{equation}
This equation implies that the the basic notion of equilibrium, $\displaystyle\frac{\partial\left|0(t)\right\rangle}{\partial t}\approx0$, is equivalent to the maximum entropy condition. As it will be shown in the numerical analysis, the entropy grows linearly over time and the only equilibrium point is at $t=0$.  The expected value of the entropy in the state (\ref{res1}) is given by equation (\ref{timeentropy}), with $N_k(t)$ given by (\ref{timedistri}). For $\gamma t<<1$, the distribution is:
\begin{equation}
N_k(t)\approx\sinh^2(\theta_k)+\gamma t\sinh(\theta_k)\cosh(\theta_k) \ , \label{tpequeno}
\end{equation}
which corresponds to a thermal distribution plus a time correction. In this limit, the entropy is given by
\begin{eqnarray}
S&=&S_{thermal}+\gamma t\beta\sum\frac{ne^{\frac{\beta n}{2}}}{e^{\beta n}-1} \nonumber \\
&=&S_{thermal}+\gamma \pi^2 t \ , \label{Sthermal}
\end{eqnarray}
where $S_{thermal}$ is the equilibrium thermodynamic entropy and the sum was calculated at the continuous limit. Note that the time-dependent term does not depend on the initial equilibrium temperature. The lattice calculations will confirm these result beyond the small time approximation.

Let us now show that the state (\ref{res1}) is in fact a non-equilibrium TFD state, in the sense of the LvN approach. The time-dependent thermal Bogoliubov transformation, based on the state (\ref{res1}) and the distribution ($\ref{timedis}$),  generates
\be
A_n(\beta,t)= \cosh(\theta+\frac{\gamma t}{2}) A_n -\sinh(\theta_n+\frac{\gamma t}{2})B_n^{\dagger} \ .  \label{tdbogo}
\ee 
such that
\be
A_n(\beta,t)\left|0(t)\right\rangle=0
\ee

The time dependent Bogoluibov transformation (\ref{tdbogo}) can be written as
\be
A_n(\beta,t)&=&\cosh(\theta_n)\left[ A_n\cosh\frac{\gamma t}{2} - B_n^{\dagger}\sinh\frac{\gamma t}{2}\right] \nonumber \\
&&-\sinh(\theta_n)\left[B_n^\dagger\cosh\frac{\gamma t}{2} + A_n \sinh\frac{\gamma t}{2}\right] \ .
\ee
Note that, comparing it with (\ref{abt}), we have
\be
A_n(t)&=&\cosh\left(\frac{\gamma t}{2}\right)A_n -\sinh\left(\frac{\gamma t}{2}\right)B_n^{\dagger} \ , \nonumber\\
B_n^{\dagger}(t)&=&\cosh\left(\frac{\gamma t}{2}\right)B_n^{\dagger}+\sinh\left(\frac{\gamma t}{2}\right)A_n \ .
\ee
These operators satisfy Eq.~(\ref{LN}) for the Hamiltonian that generated the time evolution of the squeezed vacuum $\left(H=-\frac{i\gamma}{2}\left[ab-a^{\dagger}b^{\dagger}\right]\right)$. 

Let us finish this section by comparing again the state (\ref{res1}) with the so-called time-dependent TFD vacuum  defined in (\ref{dualTFDT}), which has the following form in the free limit(disregarding a phase).
\begin {equation}
\left|TFD(t)\right\rangle=\left|0(\beta)\right\rangle=\frac{1}{Z^{1/2}}\prod_{k}\exp\left[e^{-E_k\beta/2+iE_kt}A^{\dagger}_{-k}B^{\dagger}_{k}\right]|0\rangle_A\otimes|0\rangle_B \ . \label{TTFD}
\end{equation}
This state was study in detail in \cite{myers}.  Now, following the TFD algorithm, the expected value of the number operator is given by:
\begin{equation}
\langle TFD(t)|A_k^{\dagger}A_k|TFD(t)\rangle=\frac{1}{e^{\beta E_k}-1} \ , \label{TTFDd}
\end{equation}
which is exactly the equilibrium distribution. Thus, as mentioned before, although this is a time-dependent state, from the point of view of the TFD formalism, this is related to equilibrium thermodynamics.  This is the main difference of this time dependent state with the state study here.

\section{Lattice Dissipative Conformal Theory} \label{LatDissCT}

In this section we are going to use a lattice regularization in the dissipative theory defined previously. The fields $\phi(x,t)$ and $\psi(x,t)$ will be defined in a space $M=R\otimes\Lambda$, with $R$ being the continuous time and $\Lambda$ a spatial dimensional lattice with spacing $\delta=\frac{L}{N}$, where $x\in[-L/2,L/2]$, $N$ is the number of sites and the field has spatial periodic conditions. The reason for considering a lattice theory came in two folds: first, it is an easy way to regularize the UV divergence; second, the lattice allows one to write the Hilbert space explicitly as a direct product of the Hilbert space at each site, $\otimes_{\Lambda}H_{site}$, and, therefore, the degrees of freedom may be decomposed in such a way that we can use the concept of extended entanglement entropy defined in (\ref{eentropy}); actually we can bipartite the Hilbert space in the same way as defined in (\ref{parti}).  

Using the notation of \cite{myers}, we define:
\begin{eqnarray}
Q_a&=&\delta\Phi(x_a), \: Q_{N+1}=Q_1 ,\: a=1,2..N,   \nonumber\\
P_a&=&\Pi_{\Phi}(x_a), \nonumber \\
\widetilde{Q}_a&=& \delta\Psi(x_a),\:  \widetilde{Q}_{N+1}=\widetilde{Q}_1, \nonumber \\
\widetilde{P}_a&=& \Pi_{\Psi}(x,a) \ .
\end{eqnarray}
The Hamiltonian (\ref{hcorrec}) becomes:
\begin{eqnarray}
H&=&\sum_{a=0}^N\frac{\delta}{2}\left[\left(P_a- \lambda\frac{ Q_a}{\delta}\right)^2+ m^ 2\frac{Q_a^2}{\delta^2}+\frac{1}{\delta^4}(Q_{a+1}-Q_{a})^2\right] \nonumber \\
&&-\sum_{a=0}^N\frac{\delta}{2}\left[\left(\widetilde{P}_a-\lambda\frac{ \widetilde{Q}_a}{\delta}\right)^2+ m^ 2\frac{\widetilde{Q}_a^2}{\delta^2}+\frac{1}{\delta^4}(\widetilde{Q}_{a+1}-\widetilde{Q}_{a})^2\right] \ . 
\end{eqnarray}
The mass term is an IR regulator; of course, we are interested in the limit $m\rightarrow0$. In the momentum space, the Hamiltonian can be written as:
\begin{eqnarray}
H&=&\sum_{k=0}^N \frac{\delta}{2}\left[\left(P_k-\gamma\frac{Q_k}{\delta}\right)\left(P_{-k}-\lambda\frac{Q_{-k}}{\delta}\right)+\frac{1}{\delta^2}\omega_k^2Q_kQ_{-k}\right] \nonumber \\ 
&&-\sum_{k=0}^N\frac{\delta}{2}\left[\left(\widetilde{P}_k-\gamma\frac{\widetilde{Q}_k}{\delta}\right)\left(\widetilde{P}_{-k}-\lambda\frac{\widetilde{Q}_{-k}}{\delta}\right)+\frac{1}{\delta^2}\omega_k^2\widetilde{Q}_k\widetilde{Q}_{-k}\right] \ ,
\end{eqnarray}
where
\begin{equation}
\omega_k^2=m^2+\frac{4}{\delta}\sin^2\left(\frac{\pi k}{N}\right) \ . \label{wdisc}
\end{equation}
The creation/annihilation operators can be defined as usual:
\begin{eqnarray}
A_k &=& \frac{\mu_k}{\sqrt{2}}\left[ Q_k+i\frac{P}{\mu_k^2}\right], \:\: A_k^ {\dagger} = \frac{\mu_k}{\sqrt{2}}\left[ Q_k-i\frac{P}{\mu_k^2}\right], \nonumber \\
B_k &=& \frac{\mu_k}{\sqrt{2}}\left[ \widetilde{Q}_k+i\frac{\widetilde{P}}{\mu_k^2}\right], \:\: B_k^ {\dagger} = \frac{\mu_k}{\sqrt{2}}\left[ \widetilde{Q}_k-i\frac{\widetilde{P}}{\mu_k^2}\right] \ ,
\end{eqnarray}
where $\mu_k=\sqrt{\omega_k/\delta}$.   

The Hamiltonian can be written in terms of the creation/annihilation operators as in  (\ref{hamilt}), replacing the frequency $\epsilon_k$ by  $\omega_k$, defined in (\ref{wdisc}). The distribution is now written in terms of the frequencies $\omega_k$:
\begin{equation}
N_\omega(t)=\frac{1}{e^{\beta\omega_k}-1}\left[\cosh(\gamma t)+ e^{\frac{\beta\omega_k}{2}}\sinh(\gamma t)\right]^2 , \label{n}
\end{equation}
and the entropy is given by:
\begin{equation}
S_A=\sum_{k=0}^N\left[(1+N_\omega(t))\ln(1+N_\omega(t))-N_\omega(t)\ln N_\omega(t)\right] \ . \label{Sphys} 
\end{equation}
Numerical analysis shows that this entropy grows linearly with time.


\section{The extended entanglement entropy for the dissipative lattice theory} \label{exenen}

In the previous sections we calculated the entropy for the dissipative theory, which is related to the reduced density matrix defined by a trace over all the system $B$. Differently from the entropy relative to the time-dependent TFD state studied in \cite{myers}, this entropy is time-dependent. Now, let us calculate the extended entanglement entropy for this dissipative lattice theory, where the trace is taken in the same way as in Eq.~(\ref{rhoAB}).

We define the extended reduced density matrix $\rho^d_{AB}$ for the state (\ref{res1}) partitioning the system as in (\ref{parti}). In terms of the lattice, on the oscillator number basis, we can use the following separation:
\beq
\{n_k\}_{k=1}^N=\{n_\m\}_{\m=1}^{p}\bigcup\{n_i\}_{i=p+1}^N \ ,
\eeq
and, taking the trace over the oscillators $n_i, \ i=p+1,...,N$, for the systems $A$ and $B$:
\begin{eqnarray} 
\rho^d_{AB}&=&\prod_{k=p+1}^N\frac{1}{\cosh(\theta_k+\gamma	t)}\sum_{\{n_k\}}\tanh^{2n_k}(\theta_k+\gamma	t)\times \nonumber\\
&&\times\prod_{\m=1}^{p}\sum_{\{n_\m\},\{m_\m\}}\tanh^{m_\m}(\theta_\m+\gamma	t)
\tanh^{n_\m}(\theta_\m+\gamma	t)\left.\left|n_\m\right\rangle\!\right\rangle\left\langle\!\left\langle m_\m\right|\right. \ , \label{roext}
\end{eqnarray}
where $\left.\left|n_\m\right\rangle\!\right\rangle=|n_\m\rangle\otimes|\tilde{n}_\m\rangle$ and we have used the notation $\tilde{n}_i$ to represent $B$ states. In order to carry out the extended entanglement entropy calculation more easily, let us define the time dependent parameter $\alpha_{\mu}$ by:
\be
\tanh^2(\theta_\m+\gamma t)&=&e^{-{\a_\m}} \ , \nonumber\\
\cosh^2(\theta_\m+\gamma t)&=&\frac{1}{1-e^{-\a_\m}} \ .
\ee
The extended entanglement entropy $S^d_{AB}$ is then given by:
\be
S^d_{AB}(t)&=&-Tr\rho^{d}_{AB}\ln\rho^{d}_{AB} \nonumber\\
&=&\prod_{\m=1}^{p}\coth\left(\frac{\a_\m}{4}\right)\sum_{\m=1}^{p}\left[\a_\m\frac{e^{-\frac{\a_\m}{2}}}{\left(1-e^{-\frac{\a_\m}{2}}\right)}-\ln(1-e^{-\a_\m})\right] \ .
\ee
In terms of the original variables: 
\begin{eqnarray}
S^d_{AB}(t)&=&-\prod_{\m=1}^{p}\frac{[1+\tanh(\theta_\m+\gamma t)]}{[1-\tanh(\theta_\m+\gamma t)]}\times \nonumber\\
&&\times\sum_{\m=1}^{p}\left[\ln[\tanh^2(\theta_\m+\gamma t)]\frac{\tanh(\theta_\m+\gamma t)}{[1-\tanh(\theta_\m+\gamma t)]}+\ln(1-\tanh^2(\theta_\m+\gamma t))\right] \ . \label{Sdfinal}
\end{eqnarray}
Numerical analysis will show that, at the limit $m\rightarrow0$ and for very small $\gamma$, the entropy grows linearly. For long times, the extended entanglement entropy oscillates. This is a typical behavior of spatial entropy of conformal theories and indicates that the entropy must saturate for some time. At $t=0$ we obtain the ordinary quantum entanglement entropy at finite temperature. 

Let us compare again the states (\ref{res1}) and (\ref{TTFD}), now using the density matrix. For the state (\ref{TTFD}), the density matrix is given by:
\begin{equation}
\rho(t)= |TFD(t)\rangle\langle TFD(t)| \ .
\end{equation}
By defining
\be
\tanh^2(\gamma_k)&=&e^{-\beta\omega_k+i\omega_kt} \ , \nonumber\\
\tanh^2(\bar{\gamma}_k)&=&e^{-\beta\omega_k-i\omega_kt} \ ,
\ee
we trace over the system $B$:
\be
\rho_A(t)&=&Tr_B|TFD(t)\rangle\langle TFD(t)| \nonumber\\
&=&\prod_{k=1}^{N}\frac{1}{\left(1-e^{-\b\omega_k}\right)}\sum_{m_k,n_k=0}^{\infty}\sum_{l_k=0}^{\infty}\tanh^n_k(\gamma_k)\tanh^m_k(\bar{\gamma}_k)\langle l_k|n_k\rangle_B \ \langle m_k|l_k\rangle_B \ |n_k\rangle\langle m_k| \nonumber\\
&=&\prod_{k=1}^{N}\frac{1}{\left(1-e^{-\b\omega_k}\right)}\sum_{n_k}\left[\tanh(\gamma)\tanh(\bar{\gamma})\right]^{n_k}|n_k\rangle\langle n_k| \nonumber\\
&=&\frac{1}{Z}\sum_{n_k=1}^{\infty}e^{\sum_{k=1}^Ne^{-\beta\omega_kn_k-\omega_0}}\left|n_1,...n_N\right\rangle\left\langle	n_1,...n_N\right| \ ,
\ee
which is the usual thermal equilibrium density matrix (confirming Eq.~(\ref{TTFDd})). However, note that the trace has been taken over the whole system $B$; it is due to this fact that the temporal dependence of density matrix disappears, as we will show next.

The temporal dependence of the reduced density matrix appears when the trace is taken over the complement of a subsystem. The same procedure used in (\ref{roext}) yields:  
\be
\rho_{AB}(t)&=& \displaystyle{\sum_{n_i}} \displaystyle{\sum_{\tilde{n}_i}}\langle n_i|\langle \tilde{n}_i|\rho(t)|n_i\rangle|\tilde{n}_i\rangle\nonumber\\ 
&=&\frac{1}{Z}\prod_{i=p+1}^{N}\sum_{\{n_i\}=0}^{\infty}\left[\tanh(\gamma_i)\tanh(\bar{\gamma_i})\right]^{n_i}\times \nonumber\\
&&\times\prod_{\m=1}^p\sum_{\{m_\m\},\{n_\m\}=0}^{\infty}\tanh^{n_\m}(\gamma_\m)\tanh^{m_\m}(\bar{\gamma_i})\left.\left|n_\m\right\rangle\!\right\rangle\left\langle\!\left\langle m_\m\right|\right. \ . \label{mdr}
\ee
The extended entanglement entropy for this state is given by:
\be
S_{AB}(t)^{TFD}&=&-Tr\rho_{AB}(t)\ln\rho_{AB}(t) \nonumber \\
&=&\prod_{\m=1}^{p}\frac{(1-e^{\b\omega_\m})}{F_{\m}(\b,t)}\times \nonumber\\
&&\times\sum_{\m=1}^{p}\left(\frac{e^{-\b\omega_\m}}{F_\m(\b,t)}\left[\b\omega_{\m}\left(-1+e^{\frac{\b\omega_\m}{2}}\cos\left(\frac{\omega_{\m}t}{2}\right)\right)+e^{\frac{\b\omega_m}{2}}\omega_{\m}t\sin\left(\frac{\omega_{\m}t}{2}\right)\right]\right. \nonumber\\
&&\left.-\ln(1-e^{-\b\omega_\m})\right) \ , \label{russ1}
\ee
where
\beq
F_{\m}(\b,t)=1+e^{-\b\omega_{\m}}\left[1-2e^{\frac{\b\omega_\m}{2}}\cos\left(\frac{\omega_{\m}t}{2}\right)\right] \ .
\eeq
The entropy is now time dependent; note that, differently from (\ref{Sdfinal}), it is written in terms of periodic functions. The oscillation of this entanglement entropy suggests that it must saturate for some time, which is confirmed by the numerical analysis. At short times, the entropy grows linearly. Thus, it has the same behavior as the spatial entropy of this state \cite{myers}.  


\section{Numerical Results} \label{NumRes}

Our lattice consists in $N=10.000$ sites, representing $N$ oscillators, each one with mass $m\rightarrow 1/L$, and $\beta\rightarrow CL$, for some number $C$. Using Eq.~(\ref{wdisc}), we obtain:       
\begin{equation}
\beta\omega_k =\left( \beta^2 m^2 + \frac{\beta^2}{\delta^2}4\sin^2(\frac{\pi k}{N})\right)^{1/2}
 \rightarrow  \left(C^2 + 4C^2N^2\sin^2(\frac{\pi k}{N})\right)^{1/2} \ .
\end{equation}
The thermodynamic entropy $S_A$ is calculated using this redefinition and Eqs.~(\ref{n}) and (\ref{Sphys}). For this purpose, we need two numerical inputs: the number of sites $N$ (fixed) and the temperature, represented by $C^{-1}$. First we fixed the values for the input parameter $C$ and varied $\gamma t$.

For $S_{AB}$ (the entanglement entropy), we have used the same numerical strategy as above for the thermodynamic entropy, using Eqs.~(\ref{thetan}) and (\ref{Sdfinal}). Since the product is divergent, we make $p$ ranging from $p_{min}=0.10N$ to $p_{max}=0.75N$, using it as a cutoff, to numerically maintain $S_{AB}<10^{62}$. With this procedure, the sum in Eq.~(\ref{Sdfinal}) range from $p_{min}$ to $p_{max}$.


\subsection{Numerical analysis of the thermodynamic entropy} \label{nuther}

In Fig.~(\ref{Figura1}), the graph for the thermodynamic entropy, we have fixed $\gamma t=0$ and varied $C$ in the range $[0.01,10.0]$, with steps of length $0.01$.
\begin{center}
\includegraphics[scale=0.5]{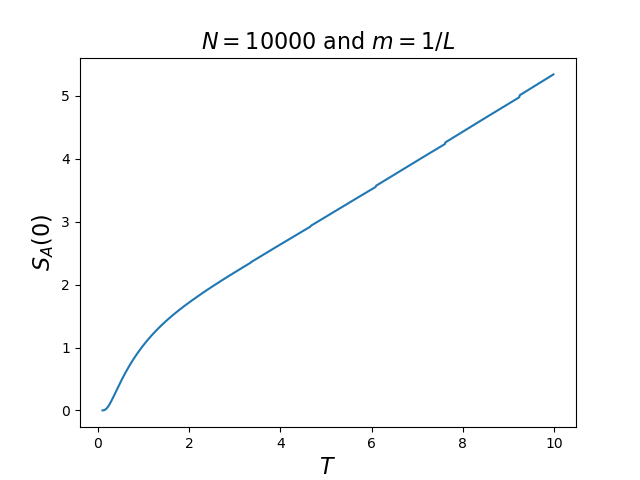}
\captionof{figure}{Plot of $S_A$ vs. the temperature $T$ of the system, which is equivalent, in our numerical analysis, to $C^{-1}$.}
\label{Figura1}
\end{center}

The time dependent entropy that comes from the trace over all the system $B$ (Fig.~(\ref{Figura2})) shows a linear behavior in all times. It has the same behavior as the entanglement entropy of conformal theories, when the Hilbert space is spatially partitioned. However, as the degrees of freedom are not confined to a given region, there is no sign of saturation. 

\begin{center}
\includegraphics[scale=0.5]{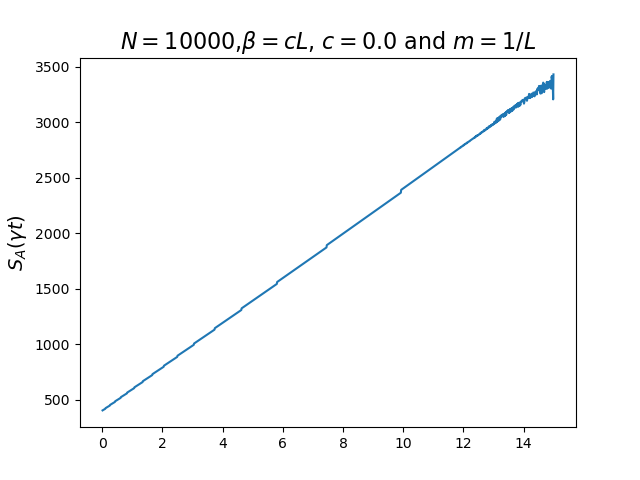}
\captionof{figure}{Linear growth of the thermodynamic entropy.}
\label{Figura2}
\end{center}

Note that at low temperatures, Fig.~(\ref{Figura6}) and Fig.~(\ref{Figura1}) show that $S^d_{AB}(0)$ varies slowly with the temperature, compared to $S_A(0)$. However, the behavior is opposite at higher temperatures, probably an effect that comes from the finite product in its very definition, remarkably in the limit $\gamma t\to 0$, when we obtain $(1-\tanh(\theta_\m+\gamma t))\to 0$ at high temperatures (since $\tanh(\theta_\m)=e^{-\frac{\beta\omega_\mu}{2}}$ ), which means that the quantum fluctuations  are higher at high temperatures. Fig.~(\ref{Figura8_T}) shows the graph of $\dot{S}_{A}=\frac{dS_A}{dt}$ vs the temperature. The graph confirms, from a minimum temperature, the independence of $S_A$ on the temperature,  which is obtained in the continuous limit and in the small $t$ approximation (Eq.~(\ref{Sthermal})). 
\begin{center}
\includegraphics[scale=0.5]{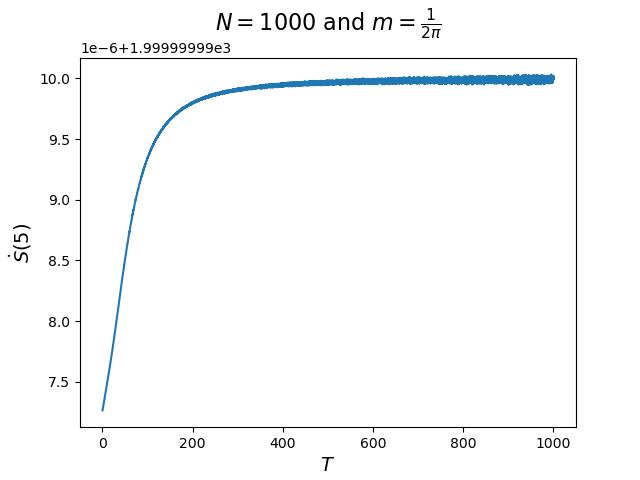}
\captionof{figure}{Plot of $\dot{S}_A$ vs. the temperature $T$ of the system.}
\label{Figura8_T}
\end{center}


\subsection{Numerical Analysis of the Extented Entanglement Entropy} \label{nuextent}
\begin{center}
\includegraphics[scale=0.5]{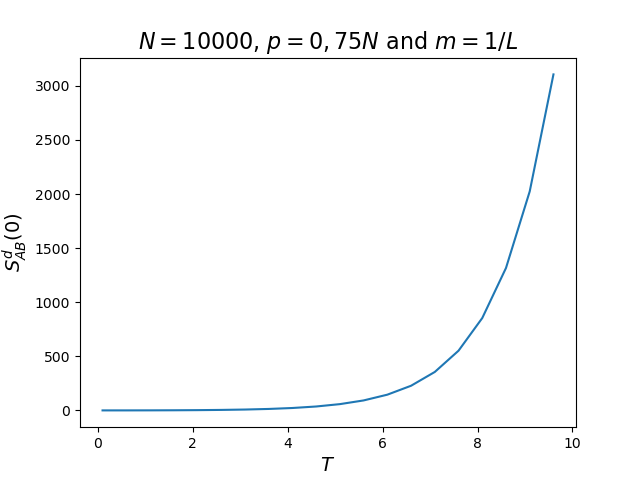}
\captionof{figure}{The behavior of $S^{d}_{AB}$ (calculated at $t=0$) as a function the temperature $T$ of the system.}
\label{Figura6}
\end{center}
In the graph for $S_{AB}^d$, we have a linear regime, showing the conformal symmetry of the system for small values of $\gamma t$, associated with $m\Rightarrow C=\beta L$ also small, as shown in Fig.~(\ref{Figura3}). Note that for short times, the behavior of the entanglement entropy $S_ {AB}^{d}$ of the dissipative system is similar to the behavior of $ S_{AB}^{TFD}$ for the state (\ref{TTFD}). The  behavior of $S_ {AB}^{d}$ (Fig.~(\ref{Figura4})), besides that it  diverges asymptotically, it is done periodically, indicating some sort of saturation. The same is true for $S_{AB}^{TFD}$, as shown in Fig.~(\ref{Figura5}). Therefore, the behavior of the extended entropy is the same as that of the spatial entanglement entropy. Note that the growth rate of $S_ {AB}^{d}$ is much higher than that of $ S_{AB}^{TFD}$. The crucial difference is the dissipative coupling between the theories. In addition to the quantum entanglement, the causal coupling between theories contributes to $S_ {AB}^d$.
\begin{center}
\includegraphics[scale=0.6]{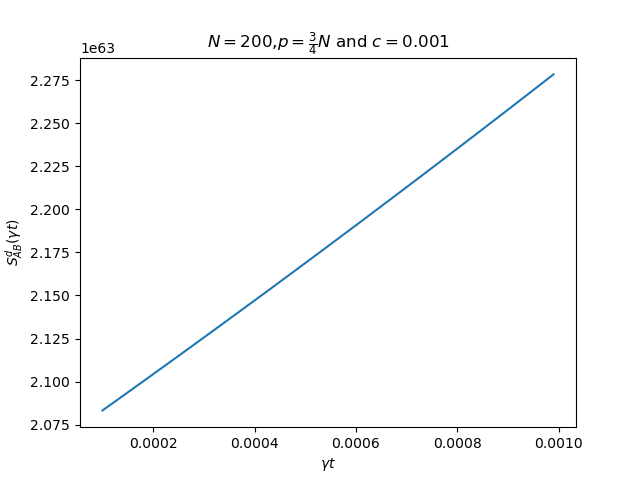}
\captionof{figure}{Plot of $S_{AB}^d$, using $N=200$ and $C=0.001$, in the regime $\gamma t\ll 1$, showing a linear behavior.}
\label{Figura3}
\end{center}
\begin{center}
\includegraphics[scale=0.5]{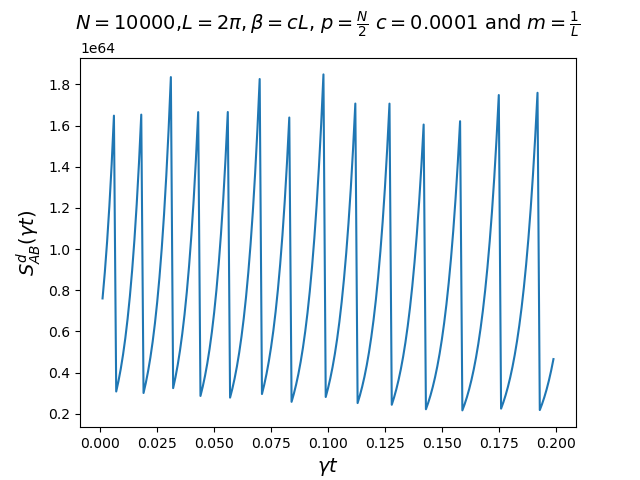}
\captionof{figure}{Behavior of $S_{AB}^d$ versus $\gamma t$.}
\label{Figura4}
\end{center}
\begin{center}
\includegraphics[scale=0.6]{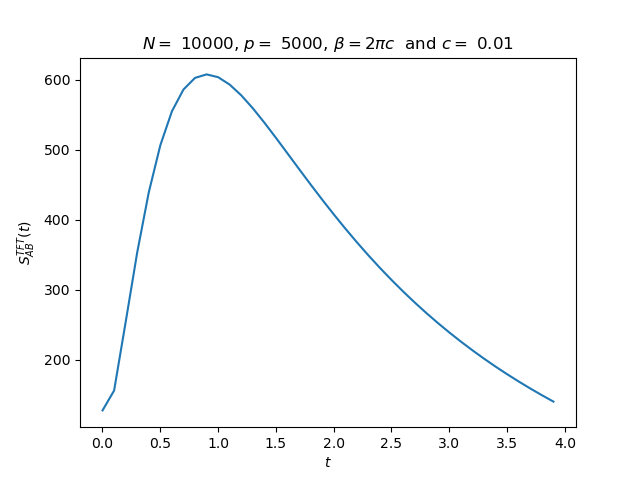}
\captionof{figure}{The behavior of $S(t)_{AB}^{TFD}$ as a function of $t$, from Eq.~(\ref{russ1}). The other parameters are shown in the top of the figure. Note the linear behavior up to approximately $t=0.4$.}
\label{Figura5}
\end{center}
	

\section{Conclusions and future directions} \label{Conc}

In this work the TFD formalism was explored, in order to study a time-dependent thermal vacuum in a dissipative  theory. This state is a consequence of a particular dissipative interaction between two theories, with respective fields $\phi$ and $\psi$, which can be interpreted as two conformal theories defined at the two asymptotic  boundaries of an AdS black hole. The initial state is prepared to be  the TFD thermal vacuum, dual to the AdS black hole.  We show that the dissipative vacuum evolves as an entangled state at finite temperature and the dissipative dynamics is controlled by the time-dependent entropy operator, defined in the non-equilibrium TFD framework. 

We use lattice field  theory techniques to calculate two types of time-dependent entropy. The first one is the expected value of the time-dependent entropy operator, which is also a time evolution controller; this is related to the entanglement between all the degrees of freedom of the two theories and it is a non-equilibrium thermodynamic entropy. Numerical analysis shows that this entropy grows linearly with time, which is typical of conformal theories. 

The second entropy calculated here is called (in the TFD literature) extended entanglement entropy: it measures the entanglement between a part of the system and its complement at finite temperature. In the TFD framework the extended entanglement entropy is calculated by defining two equal subsets of oscillators, each one belonging to a correspondent conformal theory, and tracing out the complement of each subset. Numerical analysis shows the same behavior of the spatial entanglement entropy of conformal theories. However, the extended entropy is much easier to calculate.

With regard to the TFD formalism, this work shows a time-dependent thermal state different from the state that is generally explored in the literature involving the AdS/CFT scenario, studied in \cite{malda-hartman}, and called time-dependent TFD state,  written here in Eq.~(\ref{TTFD}). The state studied here is a dissipative thermal state and it is clearly related to non-equilibrium thermodynamics. It evolves in time by the action of a Hamiltonian that has the form $\hat{H}=H_{\phi}-{H}_{\psi}+ H_{\phi\psi}$, where $H_{\phi\psi}$ is an interaction between the two systems,  while the state studied in \cite{malda-hartman} evolves by $H= H_\phi$+$H_\psi$.  The time-dependent TFD state allows one to explore various aspects involving temporal dependence of entanglement and its application in black hole physics, for a bulk geometry that is an empty space; the state studied here is more related to a Vaidya geometry. It is important to note that, as we have shown in section $V$, the expected value of the number operator in the state defined at \cite{malda-hartman} provides the bosonic thermodynamic equilibrium distribution, while in the state studied here it is a time-dependent function, which is the equilibrium distribution only at $t=0$. The extended entropy of the time-dependent TFD state is also calculated and shows the same saturation behavior of the spatial entanglement entropy of this state. 

We emphasize in this paper that the algebraic structure of TFD makes this formalism extremely suitable for applications involving the AdS/CFT scenario.  As a future work it will be also important to explore the TFD formalism to study the Eigenstate Thermalization Hypothesis in Conformal Field Theory, discussed in \cite{ETC,Tatsuhiko}. It will be also very interesting to calculate the quantum computational complexity related to the state studied here and to explore its holographic interpretation. To this end, the same covariant matrix approach used in \cite{myers} can be used in the scenario presented here.


\begin{acknowledgments}
The authors would like to thank M. B. Cantcheff, A. L. Gadelha and D. F. Z. Marchioro for useful discussions.
\end{acknowledgments}



\begin{thebibliography}{99}
\bibitem{Kitaev}
A.~Kitaev and J.~Preskill, 
Phys.\ Rev.\ Lett.\ {\bf 96} (2006) 110404, 
[arXiv:hep-th/0510092 [hep-th]].

\bibitem{Levin} 
M.~Levin and X.-G.~Wen, 
Phys.\ Rev.\ Lett.\ {\bf 96} (2006) 110405, 
[arXiv:cond-mat/0510613].

\bibitem{RT} 
S.~Ryu and T.~Takayanagi, 
Phys.\ Rev.\ Lett.\ {\bf96} (2006) 181602, 
[arXiv:hep-th/0603001].

\bibitem{masuo} 
Yoichiro Hashizume and Masuo Suzuki, 
Physica\ A {\bf 392} (2013) 17, 3518,  
[arXiv:1305.4679 [cond-mat.stat-mech]].
	
H.~Dimov, S.~Mladenov, R.~C.~Rashkov and T.~Vetsov,
Phys.\ Rev.\ D {\bf 96} (2017) 12, 126004, 
[arXiv:1705.01873 [hep-th]].

\bibitem{garvit}
P.~Garbaczewski and G.~Vitiello,
Nuovo\ Cim.\ A {\bf 44} (1978) 108.

\bibitem{ceravi}
E.~Celeghini, M.~Rasetti and G.~Vitiello,
Annals\ Phys.\  {\bf 215} (1992) 156.

\bibitem{nos}  
M.~Botta Cantcheff, A.~L.~Gadelha, D.~F.~Z.~Marchioro and D.~L.~Nedel,
Eur.\ Phys.\ J.\ C {\bf 78} (2018) 2, 105,
[arXiv:1702.02069 [hep-th]].

\bibitem{EntanglementDissipation} 
H.~Krauter, C.~A.~Muschik, K.~Jensen, W.~Wasilewski, J.~M.~Petersen, J.~I.~Cirac and E.~S.~Polzik, 
Phys.\ Rev.\ Lett. {\bf 107} (2011) 080503, 
[arXiv:1006.4344 [quant-ph]].

\bibitem{myers}  
S.~Chapman, J.~Eisert, L.~Hackl, M.~P.~Heller, R.~Jefferson, H.~Marrochio and R.~C.~Myers,
SciPost\ Phys.\ {\bf 6} (2019) 034, 
[arXiv:1810.05151 [hep-th]].

\bibitem{Balasu} 
V.~Balasubramanian, M.~B.~McDermott and M.~Van~Raamsdonk, 
Phys.\ Rev.\ D {\bf 86} (2012) 045014,
[arXiv:1108.3568 [hep-th]].

\bibitem{decoqft}
F.~Lombardo and F.~D.~Mazzitelli,
Phys.\ Rev.\ D {\bf 53} (1996) 2001,
[arXiv:hep-th/9508052].

\bibitem{inflation}
D.~Mazur and J.~S.~Heyl,
Phys.\ Rev.\ D {\bf 80} (2009) 023523,
[arXiv:0810.0521 [gr-qc]].

\bibitem{Marika} 
M.~Taylor, 
JHEP {\bf07} (2016) 040, 
[arXiv:1507.06410 [hep-th]].

\bibitem{calabrese} 
P.~Calabrese and J.~L.~Cardy, 
J.\ Stat.\ Mech.\ {\bf0406} (2004) P06002, 
[arXiv:hep-th/0405152].

\bibitem{Kim} 
S.~P.~Kim and C.~H.~Lee, 
Phys.\ Rev.\ D {\bf 62} (2000) 125020,
[arXiv:hep-ph/0005224].

\bibitem{KMMS} 
F.~C.~Khanna, A.~P.~C.~Malbouisson, J.~M.~C.~Malbouisson and A.~E.~Santana,
{\it Thermal Quantum Field Theory - Algebraic Aspects and Applications}
(World Scientific, Singapore, 2009).

\bibitem{Lewis} 
H.~R.~Lewis and W.~B.~Riesenfeld, 
J.\ Math.\ Phys.\ {\bf 10} (1969) 1458.

\bibitem{malda-hartman} 
T. Hartman and J. Maldacena, 
JHEP {\bf 1305} (2013) 014,
[arXiv:1303.1080 [hep-th]].

\bibitem{emch}
G.~G.~Emch,
{\it Alebraic Methods in Statiscal and Quantum Field Theory}
(John Wiley, New York, 1972).

\bibitem{haag}
R.~Haag,
{\it Local Quantum Physics: Fields, Particles,
Algebras}
(Springer-Verlag, New York, 1992).

\bibitem{ume2}
Y.~Takahashi and H.~Umezawa,
Coll.\ Phenomena\ {\bf 2} (1975) 55
(Reprinted in Int.\ J.\ Mod.\ Phys.\ {\bf 10} (1996) 1755).

\bibitem{ume4}
H.~Umezawa, H.~Matsumoto, M.~Tachiki,
{\it Thermofield Dynamics and Condensed States}
(North-Holland, Amsterdan, 1982).

\bibitem{rev2}
A.~Mann, M.~Revzen, H.~Umezawa and Y. Yamanaka,
Phys.\ Lett.\ A {\bf 140} (1989) 475.

\bibitem{ume1}
H.~Umezawa,
{\it Advanced Field Theory: Micro, Macro and Thermal Physics}
(AIP, New York, 1993).

\bibitem{kha2}
A.~E.~Santana and F.~C.~Khanna, 
Phys.\ Lett.\ A {\bf 203} (1995) 68.

\bibitem{kha3}
A.~E.~Santana, F.~C.~Khanna, H.~Chu and C.~Chang,
Annals\ Phys.\ {\bf 249} (1996) 481.

\bibitem{oji} 
I.~Ojima,
Annals\ Phys.\ {\bf 137} (1981) 1.

\bibitem{lands} 
N.~P.~Landsman and C.~G.~van Weert,
Phys.\ Rept.\ {\bf 145} (1987) 141.

\bibitem{Nakagawa}
K.~Nakagawa, 
PTEP\ {\bf 2015} (2015) 2, 021A01,
[arXiv:1410.6988 [cond-mat.stat-mech]].

\bibitem{Dimov} 
H.~Dimov, S.~Mladenov, R.~C.~Rashkov and T.~Vetsov, 
Phys.\ Rev.\ D {\bf 96} (2017) 12, 126004, 
[arXiv:1705.01873 [hep-th]].

\bibitem{umetime}   
H.~Umezawa and Y.~Yamanaka,
Mod.\ Phys.\ Lett.\ A {\bf 7} (1992) 3509.

\bibitem{Maldacena:2013xja}
J.~Maldacena and L.~Susskind,
Fortsch. Phys. \textbf{61} (2013), 781-811
doi:10.1002/prop.201300020
[arXiv:1306.0533 [hep-th]]

\bibitem{worm} 
P.~Gao, D.~L.~Jafferis and A.~Wall,
JHEP {\bf 12} (2017) 151, 
[arXiv:1608.05687 [hep-th]].

\bibitem{maldanovo}
J.~Maldacena, D.~Stanford and Z.~Yang,
Fortsch.\ Phys.\  {\bf 65} (2017) 5 1700034,
[arXiv:1704.05333 [hep-th]].

\bibitem{parentani}  
R.~Parentani, 
PoS QG {\bf -PH} (2007) 031,
[arXiv:0709.3943 [hep-th]].

\bibitem{collapse} 
M.~Botta~Cantcheff, 
Int.\ J.\ Mod.\ Phys.\ D {\bf 21} (2012) 1242009, 
[arXiv:1205.3113 [hep-th]].

\bibitem{VR} 
M.~Van~Raamsdonk,
Gen.\ Rel.\ Grav.\ {\bf 42} (2010) 2323, 
[arXiv:1005.3035 [hep-th]].

\bibitem {ETC}  
N.~Lashkari, A.~Dymarsky and H.~Liu,
J.\ Stat.\ Mech.\ {\bf 1803} (2018) 3, 033101,
[arXiv:1610.00302 [hep-th]].
	
\bibitem{Tatsuhiko} 
T.~Shirai and T.~Mori,
Phys.\ Rev.\ E {\bf 101} (2020) 042116,
[arXiv:1812.09713 [cond-mat.stat-mech]].


\end{thebibliography}
\end{document}